\documentclass[%
 aip,jmp,amsmath,amssymb,reprint
]{revtex4-2}

\usepackage{datetime}
\usepackage[version=4]{mhchem}
\usepackage{times,mathptmx}
\usepackage{graphicx}
\usepackage{wrapfig}
\usepackage{dcolumn}
\usepackage{eucal}
\usepackage{relsize}
\usepackage{physics}
\usepackage{xcolor}
\usepackage{hyperref}
\usepackage{siunitx}
\hypersetup{%
  colorlinks=true,
  linkcolor= [rgb]{0,0.18,0.65},
  linkbordercolor=black,
  citecolor=[rgb]{0,0.18,0.65},
  urlcolor = [rgb]{0,0.18,0.65}
  }
\usepackage{letltxmacro}
\LetLtxMacro{\OldSqrt}{\sqrt}
\newcommand{\ClosedSqrt}[1][\hphantom{3}]
{\def\DHLindex{#1}\mathpalette\DHLhksqrt}
\makeatletter
    \newcommand*\bold@name{bold}
    \def\DHLhksqrt#1#2{%
        \setbox0=\hbox{$#1\OldSqrt{#2\,}$}\dimen0=\ht0\relax%
        \advance\dimen0-0.2\ht0\relax
        \setbox2=\hbox{\vrule height\ht0 depth -\dimen0}%
        {\hbox{$#1\expandafter\OldSqrt\expandafter
        [\DHLindex]{#2\,}$}
        \lower\ifx\math@version\bold@name0.6pt\else0.4pt\fi\box2}
    }
    \renewcommand*{\sqrt}[2]
    [\ ]{\ClosedSqrt[\leftroot{-2}\uproot{1}#1]{#2}\kern0.1em} 
\makeatother
\definecolor{YKB}{rgb}{0.00,0.18,0.65}

\definecolor{beguni}{cmyk}{0.0, 0.8, 0.1, 0.1}

\def\ie{{\it i.e.},~}

\def\etal{{\it et al.}}
\def\r{\right}
\def\l{\left}
\def\lang{\left\langle}
\def\rang{\right\rangle}
\def\half{{\textstyle\frac{1}{2}}}

\def\2D{\mathsf{2D}}

\def\S#1{\mathsf{S}_\mathrm{#1}}
\def\g#1{g^\mathrm{[#1]}}
\def\dS#1{\Delta\mathsf{S}_\mathsmaller{\mathrm{#1}}}
\def\kb{\mathsf{k}_\mathsmaller{\mathrm{B}}}
\def\kapT{\kappa_\mathsmaller{\mathrm{T}}}
\def\rb{\mathbf{r}}
\def\Tstar{T^{\star}}
\def\rstar{r^{\star}}
\def\Tf{T_\mathsmaller{\mathrm{f}}}

\begin{document}

\title[]{Positional information as a universal predictor of freezing}

\author{Tamoghna Das}
\email{tamoghna.4119@gmail.com}
\affiliation{%
Center for Soft and Living Matter,\\ Institute for Basic Science (IBS), Ulsan,\\ 44919, Republic of Korea
}%
 
\author{Tsvi Tlusty}%
\email{tsvi.tlusty@gmail.com}
\altaffiliation[Also at ]{%
Department of Physics \& Department of Chemistry, Ulsan National Institute of Science and Technology, Ulsan, 44919, Republic of Korea
}%

\newdate{date}{22}{04}{2021}
\date{\displaydate{date}}

\begin{abstract}
Variation of positional information, measured by the two-body excess entropy $\S{2}$, is studied across the liquid-solid equilibrium transition in a simple two-dimensional system. Analysis reveals a master relation between $\S{2}$ and the freezing temperature $\Tf$, from which a scaling law is extracted: $-\S{2}\sim |\Tf - T| ^{-1/3}$. Theoretical and practical implications of the observed universality are discussed.

\end{abstract}

\maketitle

\section*{Introduction}
Locating thermodynamic transition lines is an essential part of material characterisation. However, broad scanning of the phase diagram of a system for any sharp changes in its thermodynamic response, primarily, heat capacity, is an exhaustive computationally-heavy procedure. As freezing of a liquid is a global structural transition, we propose here that the positional information alone can be a quick and accurate indicator of this phase transition. Such positional information of particulate systems is often directly accessible via simple X-ray scattering measurements, rendering it a practical means to efficiently chart the phase diagram. 

Liquids lack practically any {\em apriori} positional information, as all particles are arranged in a random configuration, with only occasional short range correlations. Thus, setting the totally random configuration as a reference state of {\em zero} configurational entropy, $\S{c}=0$, a liquid would have a very small negative $\S{c}$. During freezing, as spatial correlation increases, information is gained leading to further decrease in $\S{c}$. Finally, after freezing, all particles settle down in a specific crystalline symmetry described by a certain finite amount of information, corresponding to a large negative value of $\S{c}$.

In general, a disordered yet correlated arrangement of $N$ particles can be envisioned as a collection of $n$-tuples where each tuple is a possible representation of local $n$-body correlation occurring in the system with frequency $\rho^{n}$. The total configurational entropy $\S{c}$ can then be computed as an integrated contribution of such correlations as~\cite{Green:1952, Nettleton:1958, Wallace:1989} $\S{c}/N = \S{1}+\sum_{n=2}^N \S{n}$, where $\S{1}$ is the contribution of a system of non-interacting particles, namely, a classical ideal gas (up to an additive constant) and

\begin{equation}
    \S{n} = -\frac{1}{n!}~\rho^{n-1} \int d\rb^2 \g{n}
    \ln \l[\frac{\g{n}}{\mathsf{H}}\r]~.
\end{equation}
Here, $\g{n}$ is the $n$-body correlation and $\mathsf{H}$ represents the combination of all possible $(n-1)$-tuples.

For a simple liquid, large $n$-tuples can be reasonably approximated as the product of smaller $n$-tuples due to the absence of any long-lived local structures.~\cite{Kirkwood:1934}
Under such approximation,
\[
    \mathsf{H}\approx \g{3}(1,2,3)=\g{2}(1,2)\g{2}(2,3)\g{2}(3,1)
\]
and $\g{2}$ becomes the sike relevant positional correlation. The configurational entropy $\S{c}$, expressed in terms of only two-body correlation, now reads as:~\cite{Baranyai:1989}
\begin{equation}
\label{eq:exent}
\S{2} = \half + \half \rho \int d{\rb}~\l(\g{2}-1\r) 
      - \half \rho \int d{\rb}~\l(\g{2} \ln \g{2}\r)~.
\end{equation}
The positional information $-\S{2}$ is the {\em excess} information over the random configuration due to pairwise interaction in the system. Note that the upper bound of $\S{2}$ is $\half$ and not zero.

$\S{c}$, or its approximant $\S{2}$, appears to be a natural candidate to relate properties of a material to the statistical nature of its underlying internal structure. Such measures physically approximate the logarithm of the phase space volume available to the systems.~\cite{Chapman:1990} As in a classical Enskog picture, one may relate this positional information with the timescale of a particle's motion within the cage formed by its neighbours. Stemming from this idea, scaling laws~\cite{Rosenfeld:1977, Rosenfeld:1999, Dzugutov:1996} relating $\S{2}$ and transport coefficients of materials have been proposed, which have been observed to certain degrees in past and recent experiments~\cite{Vaz:2012, Ma:2013, Hansen:2018, Galloway:2020} and simulation studies.~\cite{Hoyt:2000, Chakraborty:2006, Mittal:2007, Agarwal:2011, Joy:2017, Gao:2018} In the context of glassy systems, $\S{c}$ is linked the non-Arhenius relaxation processes, as formalised in the Adam-Gibbs theory. Specifically, the relation between $\S{2}$ and glassy relaxation times was examined.~\cite{Gallo:2015, Banerjee:2017} Recently, $\S{2}$ has been recast to account for the angular correlation~\cite{Ingebrigtsen:2018} in particulate systems, and a local measure~\cite{Piaggi:2017} was formulated to study the microscopic features of materials' structure. $\S{2}$ has even been used to identify structural motifs of proteins.~\cite{Young:2007} Still, one obvious case of structural transformation, freezing, has only attracted rare attention~\cite{Giaquinta:1992, Rosenfeld:2000} in this regard.

In this study, we explore the possibility of using $\S{2}$ as an indicator of freezing transition in a simple liquid. We recognise that the two integrands contributing to Eq.(\ref{eq:exent}) originate from two independent physical processes.~\cite{Gao:2018} The first integrand originates from fluctuations in the number of particles in the grand-canonical ensemble. This positive-definite contribution, $\dS{fluc}$, termed as {\em number fluctuation entropy}, is related to the system's isothermal compressibility $\kapT$,
\begin{equation}
\dS{fluc} \equiv \half + \half \rho \int d{\rb}~\left(\g{2}-1\right) 
          =      \half \rho \kb \kapT~,
\label{eq:sfluc}
\end{equation}
where $\kb$ is the Boltzmann constant. The second integrand is a negative-definite quantity termed {\em loss entropy} as it captures the entropy lost due to pair-wise correlation that is absent from the reference ideal gas, 
\begin{equation}
\dS{loss} \equiv -\half \rho \int d{\rb}~\left(\g{2}\ln\g{2}\right)~.
\label{eq:sloss}
\end{equation}
These two entropy represent two opposing effects. On one hand, increasing spatial correlation would shrink the phase space volume resulting in more negative $\dS{loss}$. On the other hand, this would also make exchange of particles with the environment increasingly difficult, \ie $\dS{fluc}$ becomes more positive. The two-body excess entropy $\S{2}$ is governed by the interplay of these two opposing contributions.

\section*{Results}
\begin{figure*}[h!]
	\begin{center}
    \includegraphics[width=0.99\textwidth]{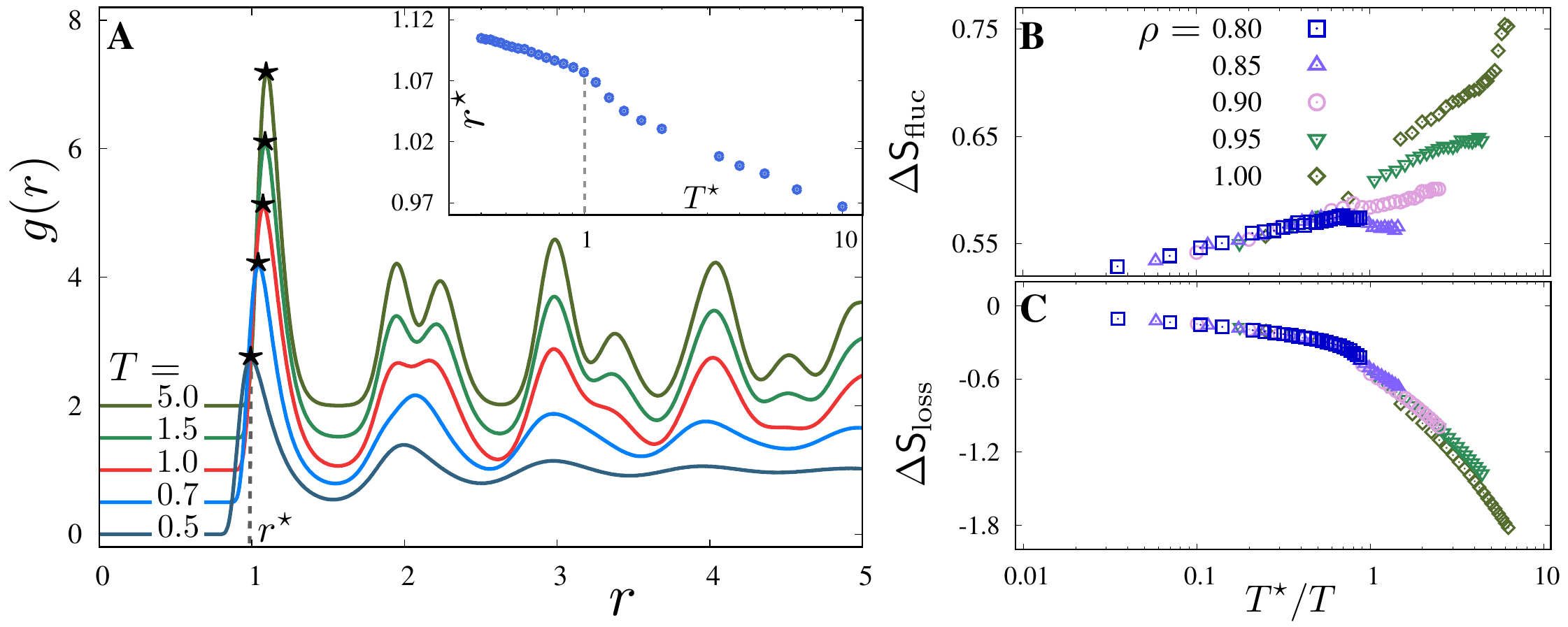}
	\end{center}
\caption{\label{fig:gofr}
(A) Pair correlation function, $g(r)$ for $\rho=0.9$ is plotted for various temperatures $T$. At $T=1.0$, a bifurcation of the second peak indicates the transition point between liquid ($T>1.0$) and solid ($T<1.0$) states. Maximal correlation for each $T$ is marked by a $\star$. Inset: position of this maximum, $\rstar$ as a function of $T$, where the transition is marked by a dashed line $T=\Tstar=1.0$. 
Two entropy terms, (B) $\dS{loss}$ and (C) $\dS{fluc}$, plotted against $\Tstar/T$ show a data collapse for  systems in the liquid phase, $T>\Tstar$, while the density dependence is evident in the solid phase, $T<\Tstar$.
}
\end{figure*}
\begin{wrapfigure}{R}{0.50\textwidth}
	\centering
    \includegraphics[width=0.45\textwidth]{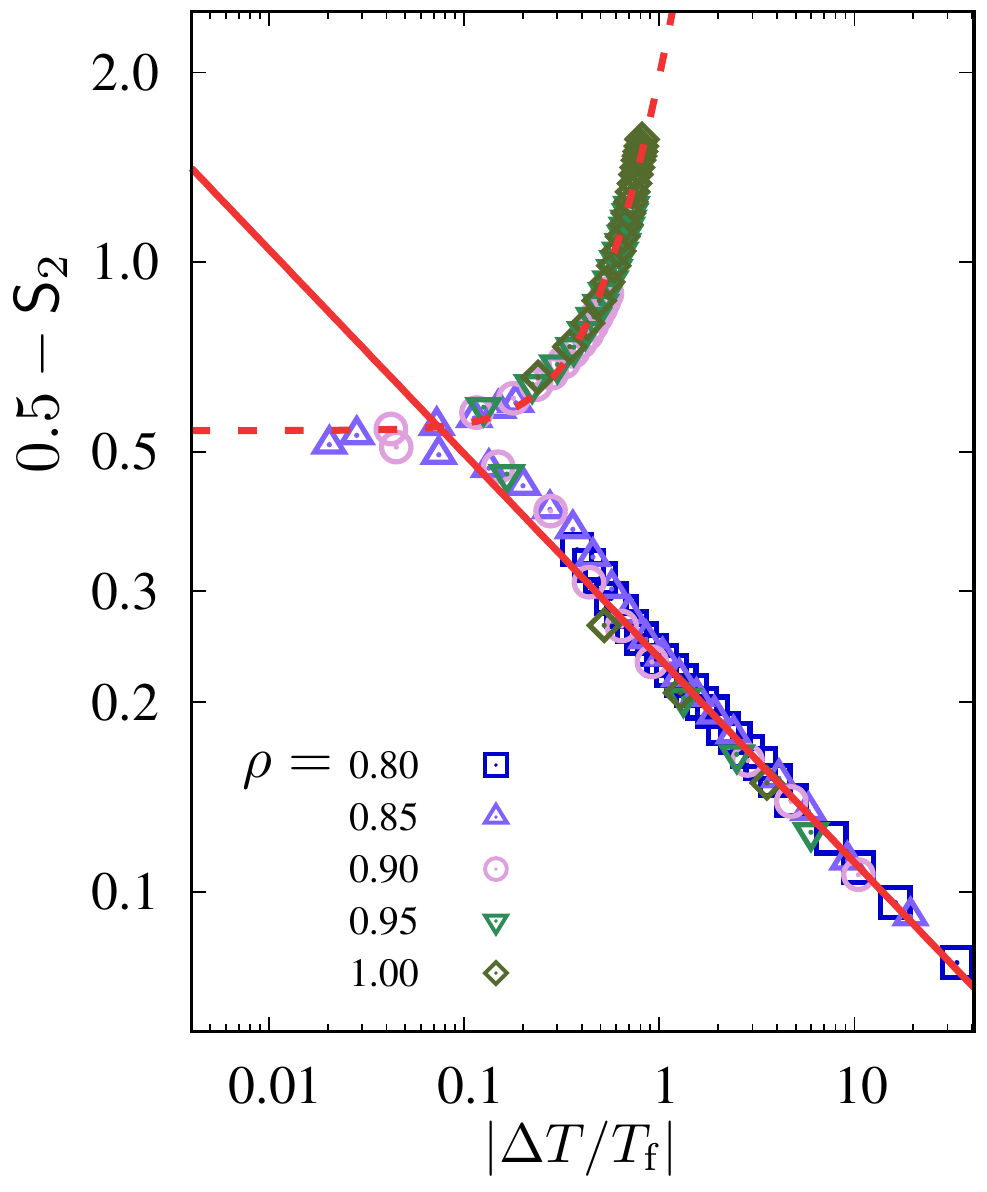}
\caption{\label{fig:scaling}
Master relation between $\S{2}$ and the freezing temperature $\Tf$. Two distinct branches of the collapsed data are fitted with power laws $x^{-\nu}$ where $x=|\Delta T/\Tf|=|\tau-1|$. The solid line, with slope $\nu\sim1/3$, corresponds to the systems in the liquid phase and the crystalline phase data are fitted with $\nu\sim-2$ is shown by (dashed line).  
}
\end{wrapfigure}
To understand the temperature dependence of structural probe $\S{2}$, we examined an ensemble of simulated equilibrium configurations of a simple system of unit-mass point particles over a wide range of densities and temperatures. The particles in this model system interact pair-wise via a Lennard-Jones potential,~\cite{LJ:1924} $\phi(r) = 4\epsilon\left[(\sigma/r)^{12}-(\sigma/r)^6\right]$ where $r=|\rb_i-\rb_j|$ is the distance between a pair $\{i,j\}$ of particles. The potential is commonly used model of simple non-polar fluids.~\cite{Vliegenthart:2000, Dunikov:2001, Charpentier:2005} The energy, length, and time scales are set by $\epsilon$, $\sigma$ and $\tau=\sqrt{\sigma^2/\epsilon}$, respectively. We have chosen to work in two-dimensions ($\2D$) for computational convenience. Equilibrium configurations of $N=\num{60000}$ particles within a periodically bounded box of size fixed by the density are simulated using canonical molecular dynamics method.~\cite{Frenkel:2006} The temperature $T$, measured in units of $\epsilon$, is fixed by a Langevin thermostat  throughout the simulation (as implemented in \texttt{LAMMPS}~\cite{Plimpton:1995}). We note that the choice of thermostat does not affect the equilibrium configuration of the system.

Since $\S{2}$, as defined in Eq.(\ref{eq:exent}), is an intensive ensemble-invariant thermodynamic property, we may conveniently replace $\g{2}$ by its canonical counterpart, the pair-correlation function, $g(r)=\rho^{-1}\lang\sum_i\delta(r-r_i)\rang$. For a typical simple liquid,~\cite{Hansen:2006} $g(r)$ shows an oscillatory behaviour at small $r$, comparable to a few particle diameter, pointing to short-range ordering, which die out exponentially with increasing $r$, as $g(r)$ tends asymptotically to unity. Upon lowering $T$ at a constant $\rho$, as crystalline order emerges in the system, the second peak of $g(r)$ splits symmetrically into two halves.~\cite{Truskett:1998} The geometric origin of this split is the $\2D$ crystalline configuration: due to the hexagonal packing of particles, one can always find two particles in the second-nearest neighbour shell positioned at angles $2\pi/3$ and $\pi$ (with respect to a suitably chosen axis).

As an example, we plot $g(r)$ of the model system (FIG.\ref{fig:gofr}A) for a set of $T$ at $\rho=0.9$. A symmetric splitting of the second peak can be observed at $T=1.0$. We also note that the position of the first peak, $\rstar$, gradually shifts towards larger values as the system condenses upon lowering $T$.~(FIG.\ref{fig:gofr}A~{\em Inset}) A rapid change in $\rstar$ is indicative of a phase transition. The crossover temperature, marked by $\Tstar$, coincides with the split of the second peak. For $T<\Tstar$, $g(r)$ is decorated with peaks at specific positions dictated by the crystalline symmetry. As $T$ crosses $\Tstar$, $g(r)$ exhibits more and more liquid-like features. Following this observational cue, we consider $\Tstar$ as a primary estimate of the freezing temperature ($\Tf$) and identify it for each of the densities. Given the $g(r)$ data, it is now straightforward to compute $\dS{fluc}$ and $\dS{loss}$ using Eqs. (\ref{eq:sfluc}) and (\ref{eq:sloss}). In the liquid state, $T>\Tstar$, the density dependence of both entropy terms can be mostly eliminated by re-scaling the temperature as $\Tstar/T$~(FIG.\ref{fig:gofr}B \& \ref{fig:gofr}C). However, the density dependence remains evident in the solids, $T<\Tstar$. 

\begin{wrapfigure}{R}{0.50\textwidth}
	\begin{center}
    \includegraphics[width=0.4\textwidth]{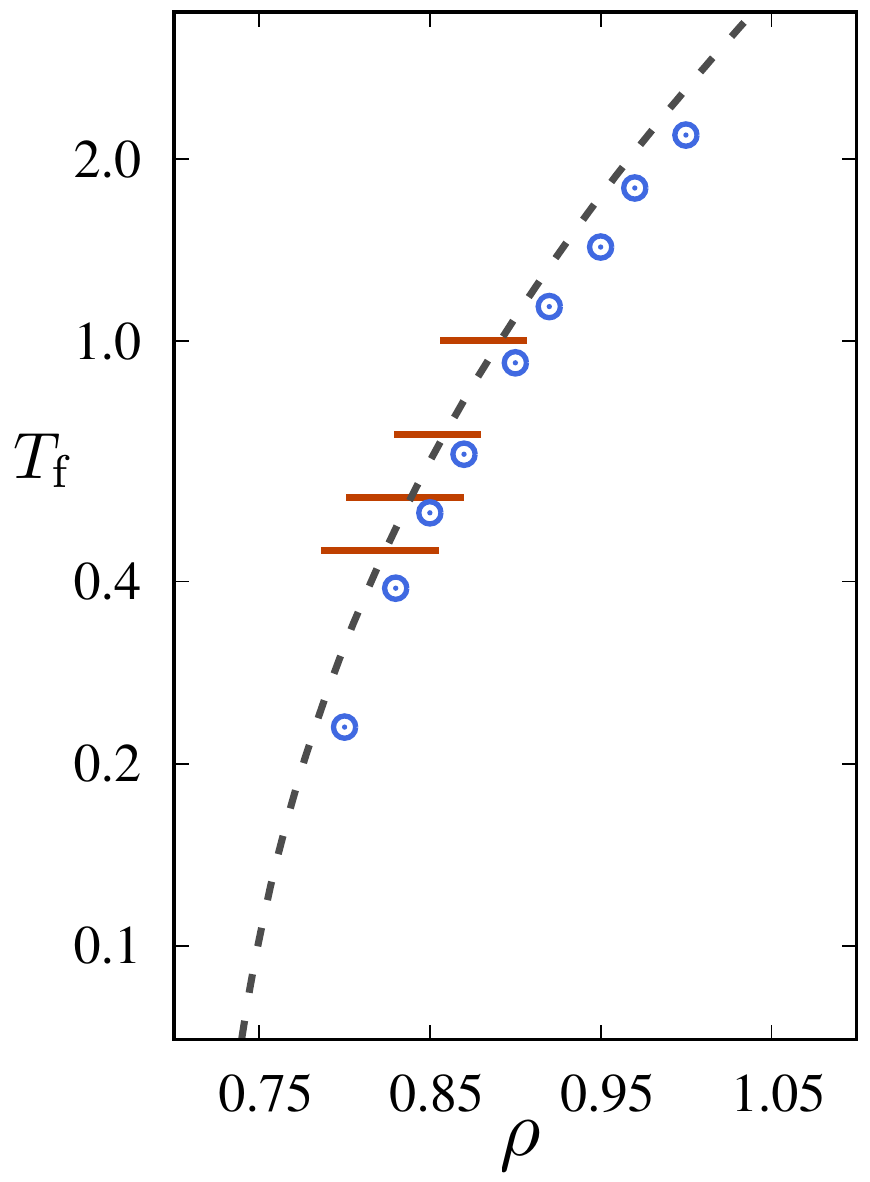}
	\end{center}
\caption{\label{fig:freez}
Density dependence of the freezing temperature, $\Tf$ (blue circles). The dashed line is the theoretical prediction for high-density Lennard-Jones system in $\2D$ ~\cite{Shi:2018}. Horizontal lines represent the only known data for the coexistence region in this system~\cite{Barker:1981}.
}
\end{wrapfigure}
Strikingly, the $\S{2}$-data for all densities collapse to a single master curve when the values of $\Tstar$ are slightly adjusted to $\Tf$. The maximal adjustment, less than $5\%$ of $\Tstar$, was required for the densest configuration ($\rho=1.0$). In FIG.\ref{fig:scaling}, we plot $-\S{2}$ shifted by $0.5$ as a function of the reduced temperature difference, $x = |\Delta T/\Tf|$ and find two distinct branches for the liquid phase (values $<0.5$) and crystalline phase (values $>0.5$). Both branches are fitted with power laws $x^{-\nu}$. A dashed line with $\nu=-2.0$ fits the crystalline branch. For the liquid branch, we find $\nu=0.326\pm0.003$ as shown by a {\em solid} line. 

The thermal dependence of $\S{2}$ closely resembles the critical behavior of heat capacity of a three-state Potts model with $\nu = 1/3$. But unlike heat capacity, which is a thermodynamic response of the system, $\S{2}$ is an intensive geometric quantity. In recent years, increasing evidences suggests a physical view of liquids as a dynamic multi-state systems,~\cite{Hoang:2015, Das:2018, Krebs:2018} in contrast to the mean field description based on homogeneous local density. Specifically, a three-state geometric picture~\cite{Das:2019} appears to capture the local dynamical aspects of liquids better than the current predictions. While the present finding strengthens this view, it also questions the traditional liquid-state theory and calls for further detailed investigations.

The values of the freezing temperature $\Tf$, as plotted in FIG.\ref{fig:freez}, are found to reside within the known boundaries of the {\em coexistence} region determined by Barker~\etal~\cite{Barker:1981} (horizontal lines). Interestingly, these values closely follow a newly-derived equation of state of $\2D$ Lennard-Jones systems.~\cite{Shi:2018} The dashed line in FIG.\ref{fig:freez} marks the temperature above which the anharmonic effects at a specific $\rho$ become prominent, and the freezing temperature is expected to be below this curve. Fulfilling this expectation, $\Tf$ appears as an overall fair estimate of freezing temperature of the present model system.

\section*{Conclusion}
The phenomenological scaling relation, $-\S{2}\sim|\Tf-T|^{-1/3}$, proposes $\S{2}$ as a reliable universal predictor of the freezing transition, at least in the case of simple liquids. Higher-order correlations would  be necessary to construct similar scaling relations for more complex network forming systems.~\cite{Baranyai:1990, Molinero:2009, Cao:2011, Dhabal:2015} We expect that in systems such as carbon-based polymeric mixtures, nano-particle composites, aggregate forming colloidal and other bio-inspired materials, $\S{2}$-based relations would provide a primary guidance for transition lines. 

The results of this study come with two implications. First, from a theoretical point of view, it suggests the {\em Corresponding States Principle}~\cite{Pitzer:1939, Pitzer:1955, Pitzer:1955a, Noro:2000} --- so far used to advocate the universality in liquids --- might now be extended to the crystalline states of matter. Thus, a unified description of the states of matter might be drawn based on the system's information content. Recalling the numerous efforts to relate $\S{2}$ with the dynamical response in condensed materials,~\cite{Dyre:2018} one might also hope to gain insight regarding the links between the dynamics and thermodynamics of matter in terms of the information exchange within the system. Second, from a practical perspective, the scaling relations allow to predict the transition temperature from simple X-ray scattering measurements of positional correlation in the system at {\em any} density and temperature, at least for a simple liquid. Since identification of a phase transitions often requires an exhaustive scan of the parameter space, this scaling relation would effectively reduce that huge search space into a single measurement.

\bibliography{exent}

\begin{thebibliography}{49}%
\makeatletter
\providecommand \@ifxundefined [1]{%
 \@ifx{#1\undefined}
}%
\providecommand \@ifnum [1]{%
 \ifnum #1\expandafter \@firstoftwo
 \else \expandafter \@secondoftwo
 \fi
}%
\providecommand \@ifx [1]{%
 \ifx #1\expandafter \@firstoftwo
 \else \expandafter \@secondoftwo
 \fi
}%
\providecommand \natexlab [1]{#1}%
\providecommand \enquote  [1]{``#1''}%
\providecommand \bibnamefont  [1]{#1}%
\providecommand \bibfnamefont [1]{#1}%
\providecommand \citenamefont [1]{#1}%
\providecommand \href@noop [0]{\@secondoftwo}%
\providecommand \href [0]{\begingroup \@sanitize@url \@href}%
\providecommand \@href[1]{\@@startlink{#1}\@@href}%
\providecommand \@@href[1]{\endgroup#1\@@endlink}%
\providecommand \@sanitize@url [0]{\catcode `\\12\catcode `\$12\catcode
  `\&12\catcode `\#12\catcode `\^12\catcode `\_12\catcode `\%12\relax}%
\providecommand \@@startlink[1]{}%
\providecommand \@@endlink[0]{}%
\providecommand \url  [0]{\begingroup\@sanitize@url \@url }%
\providecommand \@url [1]{\endgroup\@href {#1}{\urlprefix }}%
\providecommand \urlprefix  [0]{URL }%
\providecommand \Eprint [0]{\href }%
\providecommand \doibase [0]{https://doi.org/}%
\providecommand \selectlanguage [0]{\@gobble}%
\providecommand \bibinfo  [0]{\@secondoftwo}%
\providecommand \bibfield  [0]{\@secondoftwo}%
\providecommand \translation [1]{[#1]}%
\providecommand \BibitemOpen [0]{}%
\providecommand \bibitemStop [0]{}%
\providecommand \bibitemNoStop [0]{.\EOS\space}%
\providecommand \EOS [0]{\spacefactor3000\relax}%
\providecommand \BibitemShut  [1]{\csname bibitem#1\endcsname}%
\let\auto@bib@innerbib\@empty
\bibitem [{\citenamefont {Green}(1952)}]{Green:1952}%
  \BibitemOpen
  \bibfield  {author} {\bibinfo {author} {\bibfnamefont {H.~S.}\ \bibnamefont
  {Green}},\ }\href@noop {} {\emph {\bibinfo {title} {The Molecular Theory of
  Fluids}}}\ (\bibinfo  {publisher} {North Holland, Amsterdam},\ \bibinfo
  {year} {1952})\BibitemShut {NoStop}%
\bibitem [{\citenamefont {Nettleton}\ and\ \citenamefont
  {Green}(1958)}]{Nettleton:1958}%
  \BibitemOpen
  \bibfield  {author} {\bibinfo {author} {\bibfnamefont {R.~E.}\ \bibnamefont
  {Nettleton}}\ and\ \bibinfo {author} {\bibfnamefont {M.~S.}\ \bibnamefont
  {Green}},\ }\bibfield  {title} {\enquote {\bibinfo {title} {Expression in
  terms of molecular distribution functions for the entropy density in an
  infinite system},}\ }\href@noop {} {\bibfield  {journal} {\bibinfo  {journal}
  {J. Chem. Phys.}\ }\textbf {\bibinfo {volume} {29}},\ \bibinfo {pages} {1365}
  (\bibinfo {year} {1958})}\BibitemShut {NoStop}%
\bibitem [{\citenamefont {Wallace}(1989)}]{Wallace:1989}%
  \BibitemOpen
  \bibfield  {author} {\bibinfo {author} {\bibfnamefont {D.~C.}\ \bibnamefont
  {Wallace}},\ }\bibfield  {title} {\enquote {\bibinfo {title} {Statistical
  theory for the entropy of a liquid},}\ }\href@noop {} {\bibfield  {journal}
  {\bibinfo  {journal} {Phys. Rev. A}\ }\textbf {\bibinfo {volume} {39}},\
  \bibinfo {pages} {4843} (\bibinfo {year} {1989})}\BibitemShut {NoStop}%
\bibitem [{\citenamefont {Kirkwood}(1934)}]{Kirkwood:1934}%
  \BibitemOpen
  \bibfield  {author} {\bibinfo {author} {\bibfnamefont {J.~G.}\ \bibnamefont
  {Kirkwood}},\ }\bibfield  {title} {\enquote {\bibinfo {title} {Quantum
  statistics of almost classical assemblies},}\ }\href@noop {} {\bibfield
  {journal} {\bibinfo  {journal} {Phys. Rev.}\ }\textbf {\bibinfo {volume}
  {45}},\ \bibinfo {pages} {116} (\bibinfo {year} {1934})}\BibitemShut
  {NoStop}%
\bibitem [{\citenamefont {Baranyai}\ and\ \citenamefont
  {Evans}(1989)}]{Baranyai:1989}%
  \BibitemOpen
  \bibfield  {author} {\bibinfo {author} {\bibfnamefont {A.}~\bibnamefont
  {Baranyai}}\ and\ \bibinfo {author} {\bibfnamefont {D.~J.}\ \bibnamefont
  {Evans}},\ }\bibfield  {title} {\enquote {\bibinfo {title} {Direct entropy
  calculation from computer simulation of liquids},}\ }\href@noop {} {\bibfield
   {journal} {\bibinfo  {journal} {Phys. Rev. A}\ }\textbf {\bibinfo {volume}
  {40}},\ \bibinfo {pages} {3817} (\bibinfo {year} {1989})}\BibitemShut
  {NoStop}%
\bibitem [{\citenamefont {Chapman}\ and\ \citenamefont
  {Cowling}(1990)}]{Chapman:1990}%
  \BibitemOpen
  \bibfield  {author} {\bibinfo {author} {\bibfnamefont {S.}~\bibnamefont
  {Chapman}}\ and\ \bibinfo {author} {\bibfnamefont {T.~G.}\ \bibnamefont
  {Cowling}},\ }\href@noop {} {\emph {\bibinfo {title} {The mathematical theory
  of non-uniform gases: an account of the kinetic theory of viscosity, thermal
  conduction and diffusion in gases}}}\ (\bibinfo  {publisher} {Cambridge
  University Press},\ \bibinfo {year} {1990})\BibitemShut {NoStop}%
\bibitem [{\citenamefont {Rosenfeld}(1977)}]{Rosenfeld:1977}%
  \BibitemOpen
  \bibfield  {author} {\bibinfo {author} {\bibfnamefont {Y.}~\bibnamefont
  {Rosenfeld}},\ }\bibfield  {title} {\enquote {\bibinfo {title} {Relation
  between the transport coefficients and the internalentropy of simple
  systems},}\ }\href@noop {} {\bibfield  {journal} {\bibinfo  {journal} {Phys.
  Rev. A}\ }\textbf {\bibinfo {volume} {15}},\ \bibinfo {pages} {2545}
  (\bibinfo {year} {1977})}\BibitemShut {NoStop}%
\bibitem [{\citenamefont {Rosenfeld}(1999)}]{Rosenfeld:1999}%
  \BibitemOpen
  \bibfield  {author} {\bibinfo {author} {\bibfnamefont {Y.}~\bibnamefont
  {Rosenfeld}},\ }\bibfield  {title} {\enquote {\bibinfo {title} {A
  quasi-universal scaling law for atomic transport in simple fluids},}\
  }\href@noop {} {\bibfield  {journal} {\bibinfo  {journal} {J. Phys.: Condens.
  Matter}\ }\textbf {\bibinfo {volume} {11}},\ \bibinfo {pages} {5415}
  (\bibinfo {year} {1999})}\BibitemShut {NoStop}%
\bibitem [{\citenamefont {Dzugutov}(1996)}]{Dzugutov:1996}%
  \BibitemOpen
  \bibfield  {author} {\bibinfo {author} {\bibfnamefont {M.}~\bibnamefont
  {Dzugutov}},\ }\bibfield  {title} {\enquote {\bibinfo {title} {A universal
  scaling law for atomic diffusion in condensed matter},}\ }\href@noop {}
  {\bibfield  {journal} {\bibinfo  {journal} {Nature}\ }\textbf {\bibinfo
  {volume} {381}},\ \bibinfo {pages} {137} (\bibinfo {year}
  {1996})}\BibitemShut {NoStop}%
\bibitem [{\citenamefont {Vaz}\ \emph {et~al.}(2012)\citenamefont {Vaz},
  \citenamefont {Magalhaes}, \citenamefont {Fernandes},\ and\ \citenamefont
  {Silva}}]{Vaz:2012}%
  \BibitemOpen
  \bibfield  {author} {\bibinfo {author} {\bibfnamefont {R.~V.}\ \bibnamefont
  {Vaz}}, \bibinfo {author} {\bibfnamefont {A.~L.}\ \bibnamefont {Magalhaes}},
  \bibinfo {author} {\bibfnamefont {D.~L.}\ \bibnamefont {Fernandes}},\ and\
  \bibinfo {author} {\bibfnamefont {C.~M.}\ \bibnamefont {Silva}},\ }\bibfield
  {title} {\enquote {\bibinfo {title} {Vaz rv, magalh{\~a}es al, fernandes dl,
  silva cm. universal correlation of self-diffusion coefficients of model and
  real fluids based on residual entropy scaling law},}\ }\href@noop {}
  {\bibfield  {journal} {\bibinfo  {journal} {Chem. Eng. Sci.}\ }\textbf
  {\bibinfo {volume} {79}},\ \bibinfo {pages} {153} (\bibinfo {year}
  {2012})}\BibitemShut {NoStop}%
\bibitem [{\citenamefont {Ma}\ \emph {et~al.}(2013)\citenamefont {Ma},
  \citenamefont {Chen}, \citenamefont {Wang}, \citenamefont {Peng},
  \citenamefont {Han},\ and\ \citenamefont {Tong}}]{Ma:2013}%
  \BibitemOpen
  \bibfield  {author} {\bibinfo {author} {\bibfnamefont {X.}~\bibnamefont
  {Ma}}, \bibinfo {author} {\bibfnamefont {W.}~\bibnamefont {Chen}}, \bibinfo
  {author} {\bibfnamefont {Z.}~\bibnamefont {Wang}}, \bibinfo {author}
  {\bibfnamefont {Y.}~\bibnamefont {Peng}}, \bibinfo {author} {\bibfnamefont
  {Y.}~\bibnamefont {Han}},\ and\ \bibinfo {author} {\bibfnamefont
  {P.}~\bibnamefont {Tong}},\ }\bibfield  {title} {\enquote {\bibinfo {title}
  {Test of the universal scaling law of diffusion in colloidal monolayers},}\
  }\href@noop {} {\bibfield  {journal} {\bibinfo  {journal} {Phys. Rev. Lett.}\
  }\textbf {\bibinfo {volume} {110}},\ \bibinfo {pages} {078302} (\bibinfo
  {year} {2013})}\BibitemShut {NoStop}%
\bibitem [{\citenamefont {Hansen}\ \emph {et~al.}(2018)\citenamefont {Hansen},
  \citenamefont {Sanz}, \citenamefont {Adrjanowicz}, \citenamefont {Frick},\
  and\ \citenamefont {Niss}}]{Hansen:2018}%
  \BibitemOpen
  \bibfield  {author} {\bibinfo {author} {\bibfnamefont {H.~W.}\ \bibnamefont
  {Hansen}}, \bibinfo {author} {\bibfnamefont {A.}~\bibnamefont {Sanz}},
  \bibinfo {author} {\bibfnamefont {K.}~\bibnamefont {Adrjanowicz}}, \bibinfo
  {author} {\bibfnamefont {B.}~\bibnamefont {Frick}},\ and\ \bibinfo {author}
  {\bibfnamefont {K.}~\bibnamefont {Niss}},\ }\bibfield  {title} {\enquote
  {\bibinfo {title} {Evidence of a one-dimensional thermodynamic phase diagram
  for simple glass-formers},}\ }\href@noop {} {\bibfield  {journal} {\bibinfo
  {journal} {Nat. commun.}\ }\textbf {\bibinfo {volume} {9}},\ \bibinfo {pages}
  {1} (\bibinfo {year} {2018})}\BibitemShut {NoStop}%
\bibitem [{\citenamefont {Galloway}\ \emph {et~al.}(2020)\citenamefont
  {Galloway}, \citenamefont {Ma}, \citenamefont {Keim}, \citenamefont
  {Jerolmack}, \citenamefont {Yodh},\ and\ \citenamefont
  {Arratia}}]{Galloway:2020}%
  \BibitemOpen
  \bibfield  {author} {\bibinfo {author} {\bibfnamefont {K.~L.}\ \bibnamefont
  {Galloway}}, \bibinfo {author} {\bibfnamefont {X.}~\bibnamefont {Ma}},
  \bibinfo {author} {\bibfnamefont {N.~C.}\ \bibnamefont {Keim}}, \bibinfo
  {author} {\bibfnamefont {D.~J.}\ \bibnamefont {Jerolmack}}, \bibinfo {author}
  {\bibfnamefont {A.~G.}\ \bibnamefont {Yodh}},\ and\ \bibinfo {author}
  {\bibfnamefont {P.~E.}\ \bibnamefont {Arratia}},\ }\bibfield  {title}
  {\enquote {\bibinfo {title} {Scaling of relaxation and excess entropy in
  plastically deformed amorphous solids},}\ }\href@noop {} {\bibfield
  {journal} {\bibinfo  {journal} {Proc Natl Acad Sci USA}\ }\textbf {\bibinfo
  {volume} {117}},\ \bibinfo {pages} {11887} (\bibinfo {year}
  {2020})}\BibitemShut {NoStop}%
\bibitem [{\citenamefont {Hoyt}, \citenamefont {Asta},\ and\ \citenamefont
  {Sadigh}(2000)}]{Hoyt:2000}%
  \BibitemOpen
  \bibfield  {author} {\bibinfo {author} {\bibfnamefont {J.~J.}\ \bibnamefont
  {Hoyt}}, \bibinfo {author} {\bibfnamefont {M.}~\bibnamefont {Asta}},\ and\
  \bibinfo {author} {\bibfnamefont {B.}~\bibnamefont {Sadigh}},\ }\bibfield
  {title} {\enquote {\bibinfo {title} {Test of the universal scaling law for
  the diffusion coefficient in liquid metals},}\ }\href@noop {} {\bibfield
  {journal} {\bibinfo  {journal} {Phys. Rev. Lett.}\ }\textbf {\bibinfo
  {volume} {85}},\ \bibinfo {pages} {594} (\bibinfo {year} {2000})}\BibitemShut
  {NoStop}%
\bibitem [{\citenamefont {Chakraborty}\ and\ \citenamefont
  {Chakravarty}(2006)}]{Chakraborty:2006}%
  \BibitemOpen
  \bibfield  {author} {\bibinfo {author} {\bibfnamefont {S.~N.}\ \bibnamefont
  {Chakraborty}}\ and\ \bibinfo {author} {\bibfnamefont {C.}~\bibnamefont
  {Chakravarty}},\ }\bibfield  {title} {\enquote {\bibinfo {title}
  {Diffusivity, excess entropy, and the potential-energy landscape of monatomic
  liquids},}\ }\href@noop {} {\bibfield  {journal} {\bibinfo  {journal} {J.
  Chem. Phys.}\ }\textbf {\bibinfo {volume} {124}},\ \bibinfo {pages} {014507.}
  (\bibinfo {year} {2006})}\BibitemShut {NoStop}%
\bibitem [{\citenamefont {Mittal}, \citenamefont {Errington},\ and\
  \citenamefont {Truskett}(2007)}]{Mittal:2007}%
  \BibitemOpen
  \bibfield  {author} {\bibinfo {author} {\bibfnamefont {J.}~\bibnamefont
  {Mittal}}, \bibinfo {author} {\bibfnamefont {J.~R.}\ \bibnamefont
  {Errington}},\ and\ \bibinfo {author} {\bibfnamefont {T.~M.}\ \bibnamefont
  {Truskett}},\ }\bibfield  {title} {\enquote {\bibinfo {title} {Relationships
  between self-diffusivity, pack-ing fraction, and excess entropy in simple
  bulk and confined fluids},}\ }\href@noop {} {\bibfield  {journal} {\bibinfo
  {journal} {J Phys Chem B}\ }\textbf {\bibinfo {volume} {111}},\ \bibinfo
  {pages} {10054} (\bibinfo {year} {(2007)})}\BibitemShut {NoStop}%
\bibitem [{\citenamefont {Agarwal}\ \emph {et~al.}(2011)\citenamefont
  {Agarwal}, \citenamefont {Singh}, \citenamefont {Sharma}, \citenamefont
  {Alam},\ and\ \citenamefont {Chakravarty}}]{Agarwal:2011}%
  \BibitemOpen
  \bibfield  {author} {\bibinfo {author} {\bibfnamefont {M.}~\bibnamefont
  {Agarwal}}, \bibinfo {author} {\bibfnamefont {M.}~\bibnamefont {Singh}},
  \bibinfo {author} {\bibfnamefont {R.}~\bibnamefont {Sharma}}, \bibinfo
  {author} {\bibfnamefont {M.~P.}\ \bibnamefont {Alam}},\ and\ \bibinfo
  {author} {\bibfnamefont {C.}~\bibnamefont {Chakravarty}},\ }\bibfield
  {title} {\enquote {\bibinfo {title} {Relationship between structure, entropy,
  and diffusivity in water and water-like liquids},}\ }\href@noop {} {\bibfield
   {journal} {\bibinfo  {journal} {J. Phys. Chem. B}\ }\textbf {\bibinfo
  {volume} {114}},\ \bibinfo {pages} {6995} (\bibinfo {year}
  {2011})}\BibitemShut {NoStop}%
\bibitem [{\citenamefont {Joy}(2017)}]{Joy:2017}%
  \BibitemOpen
  \bibfield  {author} {\bibinfo {author} {\bibfnamefont {A.}~\bibnamefont
  {Joy}},\ }\bibfield  {title} {\enquote {\bibinfo {title} {Universal scaling
  of pair-excess entropy and diffusion in yukawa liquids},}\ }\href@noop {}
  {\bibfield  {journal} {\bibinfo  {journal} {Phys. Plasmas}\ }\textbf
  {\bibinfo {volume} {24}},\ \bibinfo {pages} {010702} (\bibinfo {year}
  {2017})}\BibitemShut {NoStop}%
\bibitem [{\citenamefont {Gao}\ and\ \citenamefont {Widom}(2018)}]{Gao:2018}%
  \BibitemOpen
  \bibfield  {author} {\bibinfo {author} {\bibfnamefont {M.~C.}\ \bibnamefont
  {Gao}}\ and\ \bibinfo {author} {\bibfnamefont {M.}~\bibnamefont {Widom}},\
  }\bibfield  {title} {\enquote {\bibinfo {title} {Information entropy of
  liquid metals},}\ }\href@noop {} {\bibfield  {journal} {\bibinfo  {journal}
  {J. Phys. Chem. B}\ }\textbf {\bibinfo {volume} {122}},\ \bibinfo {pages}
  {3550} (\bibinfo {year} {2018})}\BibitemShut {NoStop}%
\bibitem [{\citenamefont {Gallo}\ and\ \citenamefont
  {Rovere}(2015)}]{Gallo:2015}%
  \BibitemOpen
  \bibfield  {author} {\bibinfo {author} {\bibfnamefont {P.}~\bibnamefont
  {Gallo}}\ and\ \bibinfo {author} {\bibfnamefont {M.}~\bibnamefont {Rovere}},\
  }\bibfield  {title} {\enquote {\bibinfo {title} {Relation between the
  two-body entropy and the relaxation time in supercooled water},}\ }\href@noop
  {} {\bibfield  {journal} {\bibinfo  {journal} {Phys. Rev. E}\ }\textbf
  {\bibinfo {volume} {91}},\ \bibinfo {pages} {012107} (\bibinfo {year}
  {2015})}\BibitemShut {NoStop}%
\bibitem [{\citenamefont {Banerjee}\ \emph {et~al.}(2017)\citenamefont
  {Banerjee}, \citenamefont {Nandi}, \citenamefont {Sastry},\ and\
  \citenamefont {Bhattacharya}}]{Banerjee:2017}%
  \BibitemOpen
  \bibfield  {author} {\bibinfo {author} {\bibfnamefont {A.}~\bibnamefont
  {Banerjee}}, \bibinfo {author} {\bibfnamefont {M.~K.}\ \bibnamefont {Nandi}},
  \bibinfo {author} {\bibfnamefont {S.}~\bibnamefont {Sastry}},\ and\ \bibinfo
  {author} {\bibfnamefont {S.~M.}\ \bibnamefont {Bhattacharya}},\ }\bibfield
  {title} {\enquote {\bibinfo {title} {Determination of onset temperature from
  the entropy for fragile to strong liquids},}\ }\href@noop {} {\bibfield
  {journal} {\bibinfo  {journal} {J. Chem. Phys.}\ }\textbf {\bibinfo {volume}
  {147}},\ \bibinfo {pages} {024504} (\bibinfo {year} {2017})}\BibitemShut
  {NoStop}%
\bibitem [{\citenamefont {Ingebrigtsen}\ and\ \citenamefont
  {Tanaka}(2018)}]{Ingebrigtsen:2018}%
  \BibitemOpen
  \bibfield  {author} {\bibinfo {author} {\bibfnamefont {T.~S.}\ \bibnamefont
  {Ingebrigtsen}}\ and\ \bibinfo {author} {\bibfnamefont {H.}~\bibnamefont
  {Tanaka}},\ }\bibfield  {title} {\enquote {\bibinfo {title} {Structural
  predictor for nonlinear sheared dynamics in simple glass-forming liquids},}\
  }\href@noop {} {\bibfield  {journal} {\bibinfo  {journal} {Proc Natl Acad Sci
  USA}\ }\textbf {\bibinfo {volume} {115}},\ \bibinfo {pages} {87} (\bibinfo
  {year} {2018})}\BibitemShut {NoStop}%
\bibitem [{\citenamefont {Piaggi}\ and\ \citenamefont
  {Parinello}(2017)}]{Piaggi:2017}%
  \BibitemOpen
  \bibfield  {author} {\bibinfo {author} {\bibfnamefont {P.~M.}\ \bibnamefont
  {Piaggi}}\ and\ \bibinfo {author} {\bibfnamefont {M.}~\bibnamefont
  {Parinello}},\ }\bibfield  {title} {\enquote {\bibinfo {title} {Entropy based
  fingerprint for local crystalline order},}\ }\href@noop {} {\bibfield
  {journal} {\bibinfo  {journal} {J. Chem. Phys.}\ }\textbf {\bibinfo {volume}
  {142}},\ \bibinfo {pages} {114112} (\bibinfo {year} {2017})}\BibitemShut
  {NoStop}%
\bibitem [{\citenamefont {Young}\ \emph {et~al.}(2007)\citenamefont {Young},
  \citenamefont {Abel}, \citenamefont {Kim}, \citenamefont {Berne},\ and\
  \citenamefont {Friesner}}]{Young:2007}%
  \BibitemOpen
  \bibfield  {author} {\bibinfo {author} {\bibfnamefont {T.}~\bibnamefont
  {Young}}, \bibinfo {author} {\bibfnamefont {R.}~\bibnamefont {Abel}},
  \bibinfo {author} {\bibfnamefont {B.}~\bibnamefont {Kim}}, \bibinfo {author}
  {\bibfnamefont {B.~J.}\ \bibnamefont {Berne}},\ and\ \bibinfo {author}
  {\bibfnamefont {R.~A.}\ \bibnamefont {Friesner}},\ }\bibfield  {title}
  {\enquote {\bibinfo {title} {Motifs for molecular recognition exploiting
  hydrophobic enclosure in protein--ligand binding},}\ }\href@noop {}
  {\bibfield  {journal} {\bibinfo  {journal} {Proc Natl Acad Sci USA}\ }\textbf
  {\bibinfo {volume} {104}},\ \bibinfo {pages} {808} (\bibinfo {year}
  {2007})}\BibitemShut {NoStop}%
\bibitem [{\citenamefont {Giaquinta}, \citenamefont {Giunta},\ and\
  \citenamefont {Giarritta}(1992)}]{Giaquinta:1992}%
  \BibitemOpen
  \bibfield  {author} {\bibinfo {author} {\bibfnamefont {P.~V.}\ \bibnamefont
  {Giaquinta}}, \bibinfo {author} {\bibfnamefont {G.}~\bibnamefont {Giunta}},\
  and\ \bibinfo {author} {\bibfnamefont {S.~P.}\ \bibnamefont {Giarritta}},\
  }\bibfield  {title} {\enquote {\bibinfo {title} {Entropy and the freezing of
  simple liquids},}\ }\href@noop {} {\bibfield  {journal} {\bibinfo  {journal}
  {Phys. Rev. A}\ }\textbf {\bibinfo {volume} {45}},\ \bibinfo {pages} {R6966}
  (\bibinfo {year} {1992})}\BibitemShut {NoStop}%
\bibitem [{\citenamefont {Rosenfeld}(2000)}]{Rosenfeld:2000}%
  \BibitemOpen
  \bibfield  {author} {\bibinfo {author} {\bibfnamefont {Y.}~\bibnamefont
  {Rosenfeld}},\ }\bibfield  {title} {\enquote {\bibinfo {title}
  {Excess-entropy and freezing-temperature scalings for transport coefficients:
  Self-diffusion in yukawa systems},}\ }\href@noop {} {\bibfield  {journal}
  {\bibinfo  {journal} {Phys. Rev. E}\ }\textbf {\bibinfo {volume} {62}},\
  \bibinfo {pages} {7524} (\bibinfo {year} {2000})}\BibitemShut {NoStop}%
\bibitem [{\citenamefont {Lennard-Jones}(1924)}]{LJ:1924}%
  \BibitemOpen
  \bibfield  {author} {\bibinfo {author} {\bibfnamefont {J.~E.}\ \bibnamefont
  {Lennard-Jones}},\ }\bibfield  {title} {\enquote {\bibinfo {title} {On the
  determination of molecular fields. ii. from the equation of state of a
  gas},}\ }\href@noop {} {\bibfield  {journal} {\bibinfo  {journal} {Proc. R.
  Soc. Lond. A}\ }\textbf {\bibinfo {volume} {106}},\ \bibinfo {pages} {463}
  (\bibinfo {year} {1924})}\BibitemShut {NoStop}%
\bibitem [{\citenamefont {Vliegenthart}\ and\ \citenamefont
  {Lekkerkerker}(2000)}]{Vliegenthart:2000}%
  \BibitemOpen
  \bibfield  {author} {\bibinfo {author} {\bibfnamefont {G.~A.}\ \bibnamefont
  {Vliegenthart}}\ and\ \bibinfo {author} {\bibfnamefont {H.~N.~W.}\
  \bibnamefont {Lekkerkerker}},\ }\bibfield  {title} {\enquote {\bibinfo
  {title} {Predicting the gas--liquid critical point from the second virial
  coefficient predicting the gas--liquid critical point from the second virial
  coefficient},}\ }\href@noop {} {\bibfield  {journal} {\bibinfo  {journal} {J.
  Chem. Phys.}\ }\textbf {\bibinfo {volume} {112}},\ \bibinfo {pages} {5364}
  (\bibinfo {year} {2000})}\BibitemShut {NoStop}%
\bibitem [{\citenamefont {Dunikov}, \citenamefont {Malyshenko},\ and\
  \citenamefont {Zhakhovski}(2001)}]{Dunikov:2001}%
  \BibitemOpen
  \bibfield  {author} {\bibinfo {author} {\bibfnamefont {D.~O.}\ \bibnamefont
  {Dunikov}}, \bibinfo {author} {\bibfnamefont {S.~P.}\ \bibnamefont
  {Malyshenko}},\ and\ \bibinfo {author} {\bibfnamefont {V.~V.}\ \bibnamefont
  {Zhakhovski}},\ }\bibfield  {title} {\enquote {\bibinfo {title}
  {Corresponding states law and molecular dynamics simulations of the
  lennard-jones fluid},}\ }\href@noop {} {\bibfield  {journal} {\bibinfo
  {journal} {J. Chem. Phys.}\ }\textbf {\bibinfo {volume} {115}},\ \bibinfo
  {pages} {6623} (\bibinfo {year} {2001})}\BibitemShut {NoStop}%
\bibitem [{\citenamefont {Charpentier}\ and\ \citenamefont
  {Jakse}(2005)}]{Charpentier:2005}%
  \BibitemOpen
  \bibfield  {author} {\bibinfo {author} {\bibfnamefont {I.}~\bibnamefont
  {Charpentier}}\ and\ \bibinfo {author} {\bibfnamefont {N.}~\bibnamefont
  {Jakse}},\ }\bibfield  {title} {\enquote {\bibinfo {title} {Phase diagram of
  complex fluids using an efficient integral equation method},}\ }\href@noop {}
  {\bibfield  {journal} {\bibinfo  {journal} {J. Chem. Phys.}\ }\textbf
  {\bibinfo {volume} {123}},\ \bibinfo {pages} {204910} (\bibinfo {year}
  {2005})}\BibitemShut {NoStop}%
\bibitem [{\citenamefont {Frenkel}\ and\ \citenamefont
  {Smit}(1996)}]{Frenkel:2006}%
  \BibitemOpen
  \bibfield  {author} {\bibinfo {author} {\bibfnamefont {D.}~\bibnamefont
  {Frenkel}}\ and\ \bibinfo {author} {\bibfnamefont {B.}~\bibnamefont {Smit}},\
  }\href@noop {} {\emph {\bibinfo {title} {Understanding Molecular
  Simulation}}}\ (\bibinfo  {publisher} {Academic Press},\ \bibinfo {year}
  {1996})\BibitemShut {NoStop}%
\bibitem [{\citenamefont {Plimpton}(1995)}]{Plimpton:1995}%
  \BibitemOpen
  \bibfield  {author} {\bibinfo {author} {\bibfnamefont {S.}~\bibnamefont
  {Plimpton}},\ }\bibfield  {title} {\enquote {\bibinfo {title} {Fast parallel
  algorithms for short-range molecular dynamics},}\ }\href@noop {} {\bibfield
  {journal} {\bibinfo  {journal} {J. Comp. Phys.}\ }\textbf {\bibinfo {volume}
  {117}},\ \bibinfo {pages} {1} (\bibinfo {year} {1995})}\BibitemShut {NoStop}%
\bibitem [{\citenamefont {Hansen}\ and\ \citenamefont
  {McDonald}(2006)}]{Hansen:2006}%
  \BibitemOpen
  \bibfield  {author} {\bibinfo {author} {\bibfnamefont {J.-P.}\ \bibnamefont
  {Hansen}}\ and\ \bibinfo {author} {\bibfnamefont {I.~R.}\ \bibnamefont
  {McDonald}},\ }\href@noop {} {\emph {\bibinfo {title} {Theory of Simple
  Liquids}}}\ (\bibinfo  {publisher} {Academic Press},\ \bibinfo {year}
  {2006})\BibitemShut {NoStop}%
\bibitem [{\citenamefont {Truskett}\ \emph {et~al.}(1998)\citenamefont
  {Truskett}, \citenamefont {Torquato}, \citenamefont {Sastry}, \citenamefont
  {Debenedetti},\ and\ \citenamefont {Stillinger}}]{Truskett:1998}%
  \BibitemOpen
  \bibfield  {author} {\bibinfo {author} {\bibfnamefont {T.~M.}\ \bibnamefont
  {Truskett}}, \bibinfo {author} {\bibfnamefont {S.}~\bibnamefont {Torquato}},
  \bibinfo {author} {\bibfnamefont {S.}~\bibnamefont {Sastry}}, \bibinfo
  {author} {\bibfnamefont {P.~G.}\ \bibnamefont {Debenedetti}},\ and\ \bibinfo
  {author} {\bibfnamefont {F.~H.}\ \bibnamefont {Stillinger}},\ }\bibfield
  {title} {\enquote {\bibinfo {title} {Structural precursor to freezing in the
  hard-disk and hard-sphere systems},}\ }\href@noop {} {\bibfield  {journal}
  {\bibinfo  {journal} {Phys. Rev. E}\ }\textbf {\bibinfo {volume} {58}},\
  \bibinfo {pages} {3083} (\bibinfo {year} {1998})}\BibitemShut {NoStop}%
\bibitem [{\citenamefont {Shi}\ \emph {et~al.}(2018)\citenamefont {Shi},
  \citenamefont {Gu}, \citenamefont {Shen}, \citenamefont {Srivastava},
  \citenamefont {Santiso},\ and\ \citenamefont {Gubbins}}]{Shi:2018}%
  \BibitemOpen
  \bibfield  {author} {\bibinfo {author} {\bibfnamefont {K.}~\bibnamefont
  {Shi}}, \bibinfo {author} {\bibfnamefont {K.}~\bibnamefont {Gu}}, \bibinfo
  {author} {\bibfnamefont {Y.}~\bibnamefont {Shen}}, \bibinfo {author}
  {\bibfnamefont {D.}~\bibnamefont {Srivastava}}, \bibinfo {author}
  {\bibfnamefont {E.~E.}\ \bibnamefont {Santiso}},\ and\ \bibinfo {author}
  {\bibfnamefont {K.~E.}\ \bibnamefont {Gubbins}},\ }\bibfield  {title}
  {\enquote {\bibinfo {title} {High-density equation of state for a
  two-dimensional lennard-jones solid},}\ }\href@noop {} {\bibfield  {journal}
  {\bibinfo  {journal} {J. Chem. Phys.}\ }\textbf {\bibinfo {volume} {148}},\
  \bibinfo {pages} {174505} (\bibinfo {year} {2018})}\BibitemShut {NoStop}%
\bibitem [{\citenamefont {Barker}, \citenamefont {Henderson},\ and\
  \citenamefont {Abraham}(1981)}]{Barker:1981}%
  \BibitemOpen
  \bibfield  {author} {\bibinfo {author} {\bibfnamefont {J.~A.}\ \bibnamefont
  {Barker}}, \bibinfo {author} {\bibfnamefont {D.}~\bibnamefont {Henderson}},\
  and\ \bibinfo {author} {\bibfnamefont {F.~F.}\ \bibnamefont {Abraham}},\
  }\bibfield  {title} {\enquote {\bibinfo {title} {Phase diagram of the
  two-dimensional lennard-jones system; evidence for first-order
  transitions},}\ }\href@noop {} {\bibfield  {journal} {\bibinfo  {journal}
  {Physica A}\ }\textbf {\bibinfo {volume} {106}},\ \bibinfo {pages} {226}
  (\bibinfo {year} {1981})}\BibitemShut {NoStop}%
\bibitem [{\citenamefont {Hoang}, \citenamefont {Teboul},\ and\ \citenamefont
  {Odagaki}(2015)}]{Hoang:2015}%
  \BibitemOpen
  \bibfield  {author} {\bibinfo {author} {\bibfnamefont {V.~V.}\ \bibnamefont
  {Hoang}}, \bibinfo {author} {\bibfnamefont {V.}~\bibnamefont {Teboul}},\ and\
  \bibinfo {author} {\bibfnamefont {T.}~\bibnamefont {Odagaki}},\ }\bibfield
  {title} {\enquote {\bibinfo {title} {New scenario of dynamical heterogeneity
  in supercooled liquidand glassy states of 2d monatomic system},}\ }\href@noop
  {} {\bibfield  {journal} {\bibinfo  {journal} {J. Phys. Chem. B}\ }\textbf
  {\bibinfo {volume} {119}} (\bibinfo {year} {2015})}\BibitemShut {NoStop}%
\bibitem [{\citenamefont {Das}\ and\ \citenamefont {Douglas}(2018)}]{Das:2018}%
  \BibitemOpen
  \bibfield  {author} {\bibinfo {author} {\bibfnamefont {T.}~\bibnamefont
  {Das}}\ and\ \bibinfo {author} {\bibfnamefont {J.~F.}\ \bibnamefont
  {Douglas}},\ }\bibfield  {title} {\enquote {\bibinfo {title} {Quantifying
  structural dynamic heterogeneity in a dense two-dimensional equilibrium
  liquid},}\ }\href@noop {} {\bibfield  {journal} {\bibinfo  {journal} {J.
  Chem. Phys.}\ }\textbf {\bibinfo {volume} {149}},\ \bibinfo {pages} {144504}
  (\bibinfo {year} {2018})}\BibitemShut {NoStop}%
\bibitem [{\citenamefont {Krebs}\ \emph {et~al.}(2018)\citenamefont {Krebs},
  \citenamefont {Roitman}, \citenamefont {Nowack}, \citenamefont {Liepold},
  \citenamefont {Lin},\ and\ \citenamefont {Rice}}]{Krebs:2018}%
  \BibitemOpen
  \bibfield  {author} {\bibinfo {author} {\bibfnamefont {Z.}~\bibnamefont
  {Krebs}}, \bibinfo {author} {\bibfnamefont {A.~B.}\ \bibnamefont {Roitman}},
  \bibinfo {author} {\bibfnamefont {L.~M.}\ \bibnamefont {Nowack}}, \bibinfo
  {author} {\bibfnamefont {C.}~\bibnamefont {Liepold}}, \bibinfo {author}
  {\bibfnamefont {B.}~\bibnamefont {Lin}},\ and\ \bibinfo {author}
  {\bibfnamefont {S.~A.}\ \bibnamefont {Rice}},\ }\bibfield  {title} {\enquote
  {\bibinfo {title} {Transient structured fluctuations in a two-dimensional
  system with multiple ordered phases},}\ }\href@noop {} {\bibfield  {journal}
  {\bibinfo  {journal} {J. Chem. Phys.}\ }\textbf {\bibinfo {volume} {149}},\
  \bibinfo {pages} {034503} (\bibinfo {year} {2018})}\BibitemShut {NoStop}%
\bibitem [{\citenamefont {Das}\ and\ \citenamefont {Douglas}(2019)}]{Das:2019}%
  \BibitemOpen
  \bibfield  {author} {\bibinfo {author} {\bibfnamefont {T.}~\bibnamefont
  {Das}}\ and\ \bibinfo {author} {\bibfnamefont {J.~F.}\ \bibnamefont
  {Douglas}},\ }\bibfield  {title} {\enquote {\bibinfo {title} {Three-state
  heterogeneity in a model two-dimensional equilibrium liquid},}\ }\href@noop
  {} {\bibfield  {journal} {\bibinfo  {journal} {J. Mol. Liq.}\ }\textbf
  {\bibinfo {volume} {293}},\ \bibinfo {pages} {111466} (\bibinfo {year}
  {2019})}\BibitemShut {NoStop}%
\bibitem [{\citenamefont {Baranyai}\ and\ \citenamefont
  {Evans}(1990)}]{Baranyai:1990}%
  \BibitemOpen
  \bibfield  {author} {\bibinfo {author} {\bibfnamefont {A.}~\bibnamefont
  {Baranyai}}\ and\ \bibinfo {author} {\bibfnamefont {D.~J.}\ \bibnamefont
  {Evans}},\ }\bibfield  {title} {\enquote {\bibinfo {title} {Three-particle
  contribution to the configurational entropy of simple fluids},}\ }\href@noop
  {} {\bibfield  {journal} {\bibinfo  {journal} {Phys. Rev. A}\ }\textbf
  {\bibinfo {volume} {42}},\ \bibinfo {pages} {849} (\bibinfo {year}
  {1990})}\BibitemShut {NoStop}%
\bibitem [{\citenamefont {Molinero}\ and\ \citenamefont
  {Moore}(2009)}]{Molinero:2009}%
  \BibitemOpen
  \bibfield  {author} {\bibinfo {author} {\bibfnamefont {V.}~\bibnamefont
  {Molinero}}\ and\ \bibinfo {author} {\bibfnamefont {E.~B.}\ \bibnamefont
  {Moore}},\ }\bibfield  {title} {\enquote {\bibinfo {title} {Water modeled as
  an intermediate element between carbon and silicon},}\ }\href@noop {}
  {\bibfield  {journal} {\bibinfo  {journal} {J. Phys. Chem. B}\ }\textbf
  {\bibinfo {volume} {113}},\ \bibinfo {pages} {4008} (\bibinfo {year}
  {2009})}\BibitemShut {NoStop}%
\bibitem [{\citenamefont {Cao}\ \emph {et~al.}(2011)\citenamefont {Cao},
  \citenamefont {Kong}, \citenamefont {Li}, \citenamefont {Wu},\ and\
  \citenamefont {Liu}}]{Cao:2011}%
  \BibitemOpen
  \bibfield  {author} {\bibinfo {author} {\bibfnamefont {Q.-L.}\ \bibnamefont
  {Cao}}, \bibinfo {author} {\bibfnamefont {X.-S.}\ \bibnamefont {Kong}},
  \bibinfo {author} {\bibfnamefont {Y.~D.}\ \bibnamefont {Li}}, \bibinfo
  {author} {\bibfnamefont {X.}~\bibnamefont {Wu}},\ and\ \bibinfo {author}
  {\bibfnamefont {C.~S.}\ \bibnamefont {Liu}},\ }\bibfield  {title} {\enquote
  {\bibinfo {title} {Revisiting scaling laws for the diffusion coefficients in
  simple melts based on the structural deviation from hard-sphere-like case},}\
  }\href@noop {} {\bibfield  {journal} {\bibinfo  {journal} {Physica B}\
  }\textbf {\bibinfo {volume} {406}},\ \bibinfo {pages} {3114} (\bibinfo {year}
  {2011})}\BibitemShut {NoStop}%
\bibitem [{\citenamefont {Dhabal}\ \emph {et~al.}(2015)\citenamefont {Dhabal},
  \citenamefont {Nguyen}, \citenamefont {Singh}, \citenamefont {Khatua},
  \citenamefont {Molinero}, \citenamefont {Bandyopadhyay},\ and\ \citenamefont
  {Chakravarty}}]{Dhabal:2015}%
  \BibitemOpen
  \bibfield  {author} {\bibinfo {author} {\bibfnamefont {D.}~\bibnamefont
  {Dhabal}}, \bibinfo {author} {\bibfnamefont {A.~H.}\ \bibnamefont {Nguyen}},
  \bibinfo {author} {\bibfnamefont {M.}~\bibnamefont {Singh}}, \bibinfo
  {author} {\bibfnamefont {P.}~\bibnamefont {Khatua}}, \bibinfo {author}
  {\bibfnamefont {V.}~\bibnamefont {Molinero}}, \bibinfo {author}
  {\bibfnamefont {S.}~\bibnamefont {Bandyopadhyay}},\ and\ \bibinfo {author}
  {\bibfnamefont {C.}~\bibnamefont {Chakravarty}},\ }\bibfield  {title}
  {\enquote {\bibinfo {title} {Excess entropy and crystallization in
  stillinger-weber and lennard-jones fluids},}\ }\href@noop {} {\bibfield
  {journal} {\bibinfo  {journal} {J. Chem. Phys.}\ }\textbf {\bibinfo {volume}
  {143}},\ \bibinfo {pages} {164512} (\bibinfo {year} {2015})}\BibitemShut
  {NoStop}%
\bibitem [{\citenamefont {Pitzer}(1939)}]{Pitzer:1939}%
  \BibitemOpen
  \bibfield  {author} {\bibinfo {author} {\bibfnamefont {K.~S.}\ \bibnamefont
  {Pitzer}},\ }\bibfield  {title} {\enquote {\bibinfo {title} {Corresponding
  states for perfect liquids},}\ }\href@noop {} {\bibfield  {journal} {\bibinfo
   {journal} {J. Chem. Phys.}\ }\textbf {\bibinfo {volume} {7}},\ \bibinfo
  {pages} {583} (\bibinfo {year} {1939})}\BibitemShut {NoStop}%
\bibitem [{\citenamefont {Pitzer}(1955)}]{Pitzer:1955}%
  \BibitemOpen
  \bibfield  {author} {\bibinfo {author} {\bibfnamefont {K.~S.}\ \bibnamefont
  {Pitzer}},\ }\bibfield  {title} {\enquote {\bibinfo {title} {The volumetric
  and thermodynamic properties of fluids. i. theoretical basis and virial
  coefficients},}\ }\href@noop {} {\bibfield  {journal} {\bibinfo  {journal}
  {J. Am. Chem. Soc.}\ }\textbf {\bibinfo {volume} {77}},\ \bibinfo {pages}
  {3427} (\bibinfo {year} {1955})}\BibitemShut {NoStop}%
\bibitem [{\citenamefont {Pitzer}\ \emph {et~al.}(1955)\citenamefont {Pitzer},
  \citenamefont {Lippmann}, \citenamefont {R.~F.~Curl}, \citenamefont
  {Huggins},\ and\ \citenamefont {Petersen}}]{Pitzer:1955a}%
  \BibitemOpen
  \bibfield  {author} {\bibinfo {author} {\bibfnamefont {K.~S.}\ \bibnamefont
  {Pitzer}}, \bibinfo {author} {\bibfnamefont {D.~Z.}\ \bibnamefont
  {Lippmann}}, \bibinfo {author} {\bibfnamefont {J.}~\bibnamefont
  {R.~F.~Curl}}, \bibinfo {author} {\bibfnamefont {C.~M.}\ \bibnamefont
  {Huggins}},\ and\ \bibinfo {author} {\bibfnamefont {D.~E.}\ \bibnamefont
  {Petersen}},\ }\bibfield  {title} {\enquote {\bibinfo {title} {The volumetric
  and thermodynamic properties of fluids. ii. compressibility factor, vapor
  pressure and entropy of vaporization},}\ }\href@noop {} {\bibfield  {journal}
  {\bibinfo  {journal} {J. Am. Chem. Soc.}\ }\textbf {\bibinfo {volume} {77}},\
  \bibinfo {pages} {3433} (\bibinfo {year} {1955})}\BibitemShut {NoStop}%
\bibitem [{\citenamefont {Noro}\ and\ \citenamefont
  {Frenkel}(2000)}]{Noro:2000}%
  \BibitemOpen
  \bibfield  {author} {\bibinfo {author} {\bibfnamefont {M.~G.}\ \bibnamefont
  {Noro}}\ and\ \bibinfo {author} {\bibfnamefont {D.}~\bibnamefont {Frenkel}},\
  }\bibfield  {title} {\enquote {\bibinfo {title} {Extended
  corresponding-states behavior for particles with variable range
  attractions},}\ }\href@noop {} {\bibfield  {journal} {\bibinfo  {journal} {J.
  Chem. Phys.}\ }\textbf {\bibinfo {volume} {113}},\ \bibinfo {pages} {2941}
  (\bibinfo {year} {2000})}\BibitemShut {NoStop}%
\bibitem [{\citenamefont {Dyre}(2018)}]{Dyre:2018}%
  \BibitemOpen
  \bibfield  {author} {\bibinfo {author} {\bibfnamefont {J.~C.}\ \bibnamefont
  {Dyre}},\ }\bibfield  {title} {\enquote {\bibinfo {title} {Perspective:
  Excess-entropy scaling},}\ }\href@noop {} {\bibfield  {journal} {\bibinfo
  {journal} {J. Chem. Phys.}\ }\textbf {\bibinfo {volume} {149}},\ \bibinfo
  {pages} {210901} (\bibinfo {year} {2018})}\BibitemShut {NoStop}%
\end{thebibliography}%
\end{document}